\documentclass[preprint,showpacs,preprintnumbers,amsmath,amssymb]{revtex4}

\usepackage{graphicx}% Include figure files
\usepackage{bm}% bold math

\begin{document}

\title{First-exit-time probability 
density tails 
for a local height 
of a non-equilibrium Gaussian interface}

\author{G.Oshanin}
\affiliation{Laboratoire de Physique Th{\'e}orique de la Mati{\`e}re
Condens{\'e}e, Universit{\'e} Paris 6, 4 Place Jussieu, 75252 Paris,
France}

\date{\today}

\begin{abstract}
We study the long-time behavior of the 
probability density $Q_t$ of the
first exit time from a bounded interval $[-L,L]$ for 
a stochastic non-Markovian process $h(t)$ describing
fluctuations at a given point 
of a two-dimensional, infinite in both directions Gaussian interface. We show 
that $Q_t$ decays when $t \to \infty$ as a power-law $t^{-1 - \alpha}$, where $\alpha$
is non-universal and proportional to the ratio
of the thermal energy and the elastic energy of a fluctuation of size $L$. 
The fact that $\alpha$ appears to be dependent
on $L$, which is rather unusual,
implies 
that the number of 
existing moments of $Q_t$ 
depends on the size of the window $[-L,L]$. A
moment of an arbitrary order $n$, 
as a function of $L$, exists for sufficiently 
small $L$, diverges when $L$ approaches a certain threshold value $L_n$, 
and does not exist for $L > L_n$.  
For $L > L_1$, the probability density $Q_t$ is normalizable 
but does not have moments.
\end{abstract}


\maketitle

Non-equilibrium surface growth and dynamics of interfaces 
belong to an important multidisciplinary branch 
of non-equilibrium statistical mechanics,
which received much interest within several last decades
due to its
numerous practical applications. 
To name
just a few, we mention such diverse fields as crystal growth,
molecular beam epitaxy, fluctuating steps of metal surfaces,
evolution of liquid-liquid or liquid-vapor interfaces, and growing
bacterial colonies \cite{mc,zhang,krug}. 
A number of discrete
atomistic models \cite{mc,zhang,krug,doug} and stochastic
evolution equations \cite{edw,wv,kard,luse,krugg} has been proposed, 
revealing
a generic scale invariance 
as manifest in the power law behavior of
the interface width and
 correlation functions of the interface height. 
Much effort has
been also expended in understanding universal aspects of
different theoretical models and naturally occuring
processes \cite{mc,zhang,krug}.

While earlier works have focused mainly on scaling behavior, more
recent effort has been concentrated on evaluation 
of
the distribution functions of characteristic properties of
equilibrium or non-equilibrium interfaces. 
For different models of interfaces several
probability distribution functions have been determined, including,
e.g., the distribution of the width of heights in the
steady state  \cite{w}, 
maximal height distribution in one
\cite{ww1d} and two-dimensions \cite{ww2d}, 
distribution of 
spatially averaged height \cite{der} 
or of the height at any fixed
point in space \cite{pra}, and
the density of local maxima or minima of
heights \cite{min}. Last but not least,  
statistics of persistence, i.e.,
the probability that the interface height at a given point remains
persistently above (or below) its initial value during some time
interval, has been studied \cite{p,pp}.

In this paper we address a basic extreme value problem for a
non-Markovian stochastic process $h(t)$, where $h(t)$ is a  height
(measured relative to the averaged value)
at a fixed point of a
non-equilibrium two-dimensional, 
infinite in both directions
Gaussian interface, separating two phases in $d=3$. 
Using the approach of Refs.\cite{1}, 
we determine the long-time
asymptotical behavior of the "survival" probability $P_t$ that
$h(t)$ does not leave within the time interval $(0,t)$ a bounded interval $[-L,L]$,
and correspondingly, define the tail of the probability density 
$Q_t$ of the
first-exit-time $t$ from $[-L,L]$, $Q_t \sim - d P_t/dt$. 
We show that
in a striking contrast to the one-dimensional case, 
where $P_t \sim
\exp(- t^{1/2}/L^2)$ \cite{1,3}, and hence, $Q_t$ has all moments,
for a two-dimensional Gaussian interface the probability $Q_t$ is characterized by a power-law tail of
the form
\begin{equation}
\label{tail}
Q(t) \sim \frac{1}{t^{1 + \alpha}}, \;\;\; t \gg \kappa^{-1} \exp\left(\frac{\kappa L^2}{T}\right),
\end{equation}
$T$ being the temperature (measured in the units of the Boltzmann
constant $k_B$) and $\kappa$ - the interfacial tension. The exponent
$\alpha$ in Eq.(\ref{tail}) is non-universal:
\begin{equation}
\alpha = C \frac{T}{2 \kappa L^2},
\end{equation}
where $C$ is a constant, such that $\pi/16 \leq C \leq \pi/8$. The exponent
$\alpha$ is thus proportional to the ratio of the thermal energy and the
elastic energy of a fluctuation of size $L$. 
As a consequence of the $L$-dependence of $\alpha$, not all
moments of $Q_t$ exist and, what is rather unusual, the very
number of existing moments of $Q_t$ depends on the size of the
window $[-L,L]$.

Consider a two-dimensional interface whose local heights
$h_{n,m}(t)$, measured relative to the averaged value, obey an
infinite set of Langevin equations \cite{hoh}:
\begin{eqnarray}
\label{lan}  \xi \frac{d h_{n,m}(t)}{d t} = \kappa
\Big(h_{n+1,m}(t) + h_{n-1,m}(t)
+ h_{n,m+1}(t) 
+ h_{n,m-1}(t)- 4 h_{n,m}(t)\Big) +
\zeta_t^{(n,m)},
\end{eqnarray}
where $-\infty < n,m < \infty$ and $\zeta_t^{(n,m)}$ are independent
Gaussian white-noise processes:
\begin{equation}
\label{noise} \overline{\zeta_t^{(n,m)}} = 0, \;\;\;
\overline{\zeta_t^{(n,m)} \zeta_{t'}^{(n',m')}} = 2 \xi T \delta_{n,n'}
\delta_{m,m'} \delta(t - t').
\end{equation}
In Eqs.(\ref{lan}) and (\ref{noise}), $\xi$ is the friction
coefficient, 
the bar denotes averaging over thermal histories, $\delta_{n,n'}$ is
the Kronecker-delta symbol and $\delta(t)$ - the delta-function. In what follows
we set, for simplicity, $\xi = 1$, such that
the appropriate dimensionless "time" variable will be $\kappa t$. 
Dependence on $\xi$ can be restored in our final 
results by a mere replacement $T \to T/\xi$ and $\kappa \to \kappa/\xi$.

Note that Eqs.(\ref{lan}) describe, in particular, the time evolution of 
a spatially discretized Edwards-Wilkinson interface \cite{edw}, as well as  model $A$ Langevin 
dynamics \cite{hoh} of the Weeks columnar model \cite{weeks} or of a coarse-grained interface in a 
three-dimensional Ising model above the roughening temperature \cite{doug2}.

We suppose that initially the interface is flat and $h_{n,m}(t=0) =
0$ for all $n$ and $m$. Applying to Eqs.(\ref{lan}) a discrete Fourier transformation, 
solving
the resulting equation and inverting the solution, 
we find that, for
a given thermal history, the local height at the origin $h(t) =
h_{0,0}(t)$ is defined as a portfolio of an infinite number of
weighted independent Gaussian processes:
\begin{equation}
\label{kj}
h(t) = \int^t_0 d\tau e^{- 4 \kappa \tau} \sum_{n= - \infty}^{\infty}
\sum_{m = - \infty}^{\infty} I_n\left(2 \kappa \tau\right) I_m\left(2 \kappa \tau\right) \zeta_{t - \tau}^{(n,m)},
\end{equation}
where $I_n(2 \kappa\tau)$ is the modified Bessel function
of order $n$. From Eq.(\ref{kj}) one finds that at sufficiently long times \cite{doug2}:
\begin{equation}
\label{as}
\overline{h^2(t)} \sim \frac{T}{4 \pi \kappa} \ln\left(\kappa t\right).
\end{equation}

Our aim is now to define, for a non-Markovian process $h(t)$ in Eq.(\ref{kj}), 
the probability $P_t$ that $h(t)$ has not ever crossed the boundaries of the interval 
$[-L,L]$ within the time interval $(0,t)$
given the initial condition $h(t=0) = 0$, 
i.e., $P_t = {\rm Prob}({\rm max}|h(t)| < L|h(0) = 0)$, where ${\rm max}|h(t)|$ is the largest absolute value achieved by $h(t)$ within the time interval $(0,t)$. Once $P_t$ is determined, we will get an access to the behavior of 
another important probability -    
the first-exit-time probability density 
$Q_t = {\rm Prob}(t' > t| h(0)=0)$, where $t' = {\rm min}\{\tau | h(\tau) = \pm L\}$ is the time when $h(t)$ first hits either of the boundaries; hence, $Q_t$ is defined as
\begin{equation}
Q_t dt \equiv - \frac{d P_t}{dt} dt = {\rm Prob}\left(t < t' \leq t + dt| h(0) = 0\right).
\end{equation}

Focusing on the large-$t$ asymptotical behavior,
we note that it is not really
important
how we define
$\zeta_t^{(n,m)}$ - as continuous in time functions or as discrete
processes, provided that we keep all essential features of noise.
We thus divide, at fixed $t$, the interval $(0,t)$ into $N$ 
 small subintervals $\Delta$, (such that $\Delta N \equiv t$),
and assume that
\begin{equation}
\zeta_{\tau}^{(n,m)} = \left(\frac{2 T}{\Delta}\right)^{1/2} \, S_{[\tau/\Delta]}^{(n,m)},
\end{equation}
where $[x]$ denotes the floor function. In other words,
we assume that $\zeta_{\tau}^{(n,m)}$ is constant and equal to
$\sqrt{2 T/\Delta} \, S_k^{(n,m)}$ within the $k$-th subinterval, $k=0,1,
\ldots, N-1$, where $\{S_k^{(n,m)}\}$ is an infinite set
of independent identically distributed
random variables with normal distribution $N[0,1]$.

Then, $h(t)$ can be expressed as a weighted
sum of an infinite number of independent discrete 
noise processes:
\begin{eqnarray}
\label{def2} h(N) = \sum_{n = - \infty}^{\infty} \sum_{n = - \infty}^{\infty}  \sum_{l=1}^{N} \sigma^{(n,m)}_{l} S^{(n,m)}_{N - l},
\end{eqnarray}
with weights
\begin{equation}
\sigma^{(n,m)}_{l} = \left(\frac{T}{2 \Delta \kappa^2}\right)^{1/2} \int_{2 \kappa \Delta (l -
1)}^{2 \kappa \Delta l} du \, e^{- 2 u} \, I_n(u) I_m(u).
\end{equation}
Squaring Eq.(\ref{def2}) and
averaging the resulting expression with respect to the distribution
of i.i.d. variables $\{S_k^{(n,m)}\}$, we get
\begin{equation}
\label{msd2} \overline{h^2(N)} = \sum_{l=1}^N
\tilde{\sigma}^2_l,
\end{equation}
which expression
 introduces an effective variance $\tilde{\sigma}^2_l$, defined as
\begin{eqnarray}
\label{var} \tilde{\sigma}^2_l = \sum_{n = - \infty}^{\infty} \sum_{m = - \infty}^{\infty}
\left(\sigma^{(n,m)}_l\right)^2 
= \frac{T}{2 \Delta \kappa^2} \int^{2 \kappa \Delta
l}_{2 \kappa \Delta (l-1)} du_1 \int^{2 \kappa \Delta l}_{2 \kappa \Delta (l-1)} du_2 \, f(u_1,u_2),
\end{eqnarray}
where
\begin{equation}
f(u_1,u_2) = e^{-2 u_1 -
2 u_2} \, I_0^2\left(u_1 + u_2\right).
\end{equation}
Noticing that $\exp(- 2 x) I_0^2(x)$ is a monotonically decreasing function of $x$, one finds that $\tilde{\sigma}^2_l$ is bounded by two monotonically decreasing functions of $l$:
\begin{equation}
\label{in}
\exp\left(- 8 \kappa \Delta l\right) I_0^2\left(4 \kappa \Delta l\right) \leq \frac{\tilde{\sigma}^2_l}{2 \Delta T} \leq \exp\left(- 8 \kappa \Delta (l - 1)\right) I_0^2\left(4 \kappa \Delta (l - 1)\right).
\end{equation}
When $l \gg 1$, these bounds become very sharp and hence, with a good accuracy,
\begin{equation}
\label{nn}
\tilde{\sigma}^2_l \approx 2 \Delta T \exp\left(- 8 \kappa \Delta l\right) I_0^2\left(4 \kappa \Delta l\right).
\end{equation}
Substituting Eq.(\ref{nn}) into Eq.(\ref{msd2}), we recover the result in Eq.(\ref{as}).

Define now the following event: An $N$-step discrete-time trajectory
$h(N)$, Eq.(\ref{def2}), commencing at the origin,
does not leave the interval $[-L,L]$, or in other words,
the maximal absolute value, ${\rm max}|h(N)|$, which process $h(N)$ achieves for a given realization of noise, 
is 
less than $L$.  Probability of such an event we denote as
$P_N = {\rm Prob}({\rm max}|h(N)| < L)$.

Now, condition ${\rm max}|h(N)| < L$ implies
that the absolute value of
any ascending partial sum
\begin{equation}
h_{k}(N) = \sum_{n = - \infty}^{\infty} \sum_{m = - \infty}^{\infty} \sum_{l = N - k + 1}^{N} \sigma_l^{(n,m)} S_{N - l}^{(n,m)},
\end{equation}
which define the values of the local height $h(N)$
at consecutive discrete "time"
moments $k$, $k=1,2, \ldots, N$, is bounded from above by $L$.

We prefer, however, to deal 
with the \textit{descending} partial sums:
\begin{equation}
\label{aa}
h'_{k}(N) = \sum_{n = - \infty}^{\infty} \sum_{m = - \infty}^{\infty} \sum_{l = 1}^{k} \sigma_l^{(n,m)} S_{N - l}^{(n,m)},
\end{equation}
which define the trajectory $h_k(N)$ evolving  in the inverse
time $N - k$ and shifted by a constant (realization-dependent) value $h_{N}(N)$.
Clearly, 
$P_N = {\rm Prob}({\rm max}|h(N)| < L) = {\rm Prob}({\rm max}|h'(N)| < L)$.

Let ${\rm I}( {\rm max}|h'(t)| < L)$
be the indicator function:
\begin{equation}
\label{indicator1}
{\rm I}( {\rm max}|h'(N)| < L) =
\begin{cases}
1\,,~{\rm max}|h'(N)| < L\,,\\
0 \,,~{\rm max}|h'(N)| > L\,.
\end{cases}
\end{equation}
and ${\rm R}_L(x)$ - a rectangular
function, such that:
\begin{equation}
{\rm R}_L(x) = \int_{-\infty}^{\infty} \frac{d y}{\pi}
\frac{\sin(L y)}{y} e^{i y x} =
\begin{cases}
1, & ~ |x| < L,\\
1/2, & ~ x = \pm L, \\
0,  & ~ |x| > L.
\end{cases}
\label{indicator}
\end{equation}
Then, using the definition of
the descending partial sums in Eq.(\ref{aa}) and  Eq.(\ref{indicator}),
we write down Eq.(\ref{indicator1}) as the following $N$-fold integral:
\begin{eqnarray}
\label{indicator4}
{\rm I}\Big( {\rm max}|h'(N)| < L\Big) = \prod_{k=1}^N
{\rm I}\Big(|h'_{k}(N)| < L\Big) = 
 \int_{-\infty}^{\infty}
\ldots \int_{-\infty}^{\infty} \prod_{k=1}^N \frac{dy_k}{\pi}
\frac{\sin(L y_k )}{y_k} e^{i \; y_k \; h'_{k}(N)}.
\end{eqnarray}
Averaging the indicator function in Eq.(\ref{indicator4}) with respect to the
 distribution of i.i.d. random variables
$\{S_k^{(n,m)}\}$, and changing integration variables 
(see Refs.\cite{1} for more details), we find eventually the following exact 
representation of $P_N$:
\begin{eqnarray}
\label{f} P_N = \int_{-L}^{L} \ldots \int_{-L}^{L}
\prod_{k=1}^{N} \frac{dh_k}{\sqrt{2 \pi} \tilde{\sigma}_k} \;
\exp\left[- \frac{\left(h_k - h_{k-1}\right)^2}{2 \tilde{\sigma}_k^2}
\right],
\end{eqnarray}
where $h_0$ is fixed, $h_0 \equiv 0.$ Once $P_N$ is known, the desired
survival  
probability $P_t$ 
can be obtained as an appropriate limit: $P_t = \lim_{\Delta \to 0, N \to \infty} P_N$, 
with $\Delta N = t$ kept fixed. 

The $N$-fold integral in Eq.(\ref{f}) can not be, of course, evaluated
exactly and 
one has to resort  to controllable approximations.
Here, using the 
approach of Refs.\cite{1},  we construct rigorous lower 
and upper bounds on $P_N$ in Eq.(\ref{f}),  
taking an advantage of the following property of $P_N$ in Eq.(\ref{f}) \cite{1}:\\
$P_N = P_N(\tilde{\sigma}_1,\tilde{\sigma}_2,\tilde{\sigma}_3, \ldots,\tilde{\sigma}_N)$ is a monotonically decreasing function of any variable $\tilde{\sigma}_k$.\\
This signifies that replacing any or all $\tilde{\sigma}_k$ 
by $\Sigma(k)$, such that
$\tilde{\sigma}_k \leq \Sigma(k)$, we will decrease the right-hand-side of Eq.(\ref{f}) 
and arrive at the \textit{lower} bound on $P_N$; if, on
contrary, we will replace one or all $\tilde{\sigma}_k$ by
$\tilde{\Sigma}(k)$, such that 
$\tilde{\sigma}_k \geq \tilde{\Sigma}(k)$, 
we will \textit{increase} the right-hand-side of Eq.(\ref{f}) and
obtain an \textit{upper} bound on $P_N$.

We start with a lower bound on $P_N$. As we have already noticed, $\exp(- 2 x) I_0^2(x)$ in Eq.(\ref{var}) is a monotonically decreasing function of $x$. Consequently, setting in the integrand $u_2 \equiv 0$, and integrating over $d u_2$, we have 
\begin{eqnarray}
\tilde{\sigma}_k^2 \leq \Sigma(k)^2 = \frac{T}{\kappa} \int_{2 \kappa \Delta (k-1)}^{2 \kappa \Delta k} du e^{- 2 u} I_0^2(u) = t^{(upp)}_k - t^{(upp)}_{k - 1},
\end{eqnarray}
where
\begin{equation}
\label{ti}
t^{(upp)}_k \equiv \frac{T}{\kappa} \int^{2 \kappa \Delta k}_{0} du \, e^{- 2 u} I_0^2(u).
\end{equation}
Therefore, $P_N$ in Eq.(\ref{f}) is bounded from \textit{below} by
\begin{eqnarray}
\label{fl} P_N \geq \int_{-L}^{L} \ldots \int_{-L}^{L}
\prod_{k=1}^{N} \frac{dh_k}{\sqrt{2 \pi \left(t^{(upp)}_{k} - t^{(upp)}_{k - 1}\right)}} \;
\exp\left[- \frac{\left(h_k - h_{k-1}\right)^2}{2 \left(t^{(upp)}_{k} - t^{(upp)}_{k - 1}\right)}
\right],
\end{eqnarray}
where $h_0$ is fixed, $h_0 \equiv 0.$ 

One notices now that the expression on the right-hand-side of 
Eq.(\ref{fl}) describes the probability that an $N$-step trajectory of a 
Brownian motion evolving in "time" $t^{(upp)}_k$ does not leave an interval $[-L,L]$,
 which yields, in the leading order,
\begin{equation}
\label{pu}
P_N \geq \exp\left( - \frac{\pi^2}{ 8 L^2} t^{(upp)}_N\right) = \exp\left( - \frac{\pi^2 T}{ 8 \kappa L^2} \int^{2 \kappa \Delta N}_0 du \exp(- 2 u) I_0^2(u)\right).
\end{equation}
Defining the asymptotical large-$N$ behavior of the integral in the exponential in Eq.(\ref{pu}) and turning to 
the limit $\Delta \to 0$, $N \to \infty$ with $\Delta N = t$ kept fixed,
we arrive at the following lower bound on $P_t$:
\begin{equation}
\frac{\ln(P_t)}{\ln(\kappa t)} \geq - \frac{\pi}{8} \frac{T}{2 \kappa L^2}, \;\;\; \kappa t \gg \exp(\kappa L^2/T).
\end{equation} 

Consider next an \textit{upper} bound on $P_N$, Eq.(\ref{f}). Using the inequality in Eq.(\ref{in}) and an evident inequality:
$\exp(- 2 x) I_0^2(x) \geq 1/(2 \pi x + 1)$, which holds for any $x \geq 0$, we have
\begin{eqnarray}
\label{inn}
\tilde{\sigma}^2_k \geq 2 \Delta T \exp\left(- 8 \kappa \Delta k\right) I_0^2\left(4 \kappa \Delta k\right) \geq \frac{2 \Delta T}{8 \pi \kappa \Delta k + 1} \geq \tilde{\Sigma}(k)^2 =  t^{(low)}_k - t^{(low)}_{k - 1},
\end{eqnarray}
whith
\begin{equation}
t^{(low)}_k = \frac{T}{4 \pi \kappa} \ln\left(8 \pi \kappa \Delta (k + 1) + 1\right).
\end{equation}
Hence, in virtue of Eq.(\ref{inn}), $P_N$ in Eq.(\ref{f}) is bounded from \textit{above} by
\begin{eqnarray}
\label{fu} P_N \leq \int_{-L}^{L} \ldots \int_{-L}^{L}
\prod_{k=1}^{N} \frac{dh_k}{\sqrt{2 \pi \left(t^{(low)}_{k} - t^{(low)}_{k - 1}\right)}} \;
\exp\left[- \frac{\left(h_k - h_{k-1}\right)^2}{2 \left(t^{(low)}_{k} - t^{(low)}_{k - 1}\right)}
\right],
\end{eqnarray}
where $h_0$ is fixed, $h_0 \equiv 0.$ 

One recognizes now that the rhs of Eq.(\ref{fu}) is 
the probability that an $N$-step trajectory of a Brownian motion evolving in "time" $t^{(low)}_k$ does not leave an interval $[-L,L]$. 
Hence, the following \textit{upper} bound on $P_N$ is valid:
\begin{equation}
\label{upper}
P_N \leq \exp\left( - \frac{\pi^2}{ 8 L^2} t^{(low)}_N\right) \sim \exp\left( - \frac{\pi T}{ 16 \kappa L^2} \ln\left(\kappa \Delta N\right)\right),
\end{equation}
which yields the following upper bound on $P_t$:
\begin{equation}
\frac{\ln(P_t)}{\ln(\kappa t)} \leq - \frac{\pi}{16} \frac{T}{2 \kappa L^2}, \;\;\; \kappa t \gg \exp(\kappa L^2/T).
\end{equation} 
Since $P_t$ is, evidently, a monotonically decreasing function of time, we infer that the 
large-$t$ asymptotical behavior of $P_t$ is described by a power-law of the 
form $P_t \sim (\kappa t)^{-\alpha}$, where $\alpha = C T/2 \kappa L^2$ and $C$ is a constant, 
such that $\pi/16 \leq C \leq \pi/8$. Consequently, the long-time tail of the 
first-exit-time probability density $Q_t$ is a power-law, $Q_t \sim t^{-1 - \alpha}$, Eq.(\ref{tail}).

In conclusion, we have shown that the 
probability density $Q_t$ of the
first-exit-time $t$ from a bounded interval $[-L,L]$ for 
a non-Markovian process $h(t)$ describing
fluctuations at a given point 
of a two-dimensional, infinite in both directions Gaussian interface, 
is characterized by a power-law tail
in Eq.(\ref{tail}). The exponent $\alpha$ is non-universal and proportional to the ratio
of the thermal energy and the elastic energy of a fluctuation of size $L$, and thus depends on $L$.
We note that a power-law behavior of $P_t$ is not counterintuitive, and could, 
in principle, 
be expected from a heuristic estimate $P_t \sim \exp(-\overline{h^2(t)}/L^2)$ \cite{osh2} combined with a logarithmic growth
of the second moment of $h(t)$, Eq.(\ref{as}). 
On the other hand, the dependence of $\alpha$ on $L$ is a rather unusual fact 
which entails unusual behavior of the moments of $Q_t$; namely, 
the number of exisiting moments of $Q_t$ appears to be defined by the size of the window in which 
the stochastic process $h(t)$ is observed. More specifically, the number $n$ 
of existing moments depends on the relation between $L$ and a discrete
set of characteristic lengths $L_n = (C T/2 \kappa n)^{1/2}$; the
condition that $Q_t$ has exactly $n$ moments is, in fact, equivalent to the requirement that $L$ obeys the following
double-sided inequality: $L_{n+1} < L < L_n$. 
In other words, a moment of an arbitrary order $n$, 
as a function of $L$, exists for sufficiently small $L$, diverges when $L$ approaches $L_n$, 
and does not exist for $L > L_n$. For $L \geq L_1$, the probability density $Q_t$ is normalizable but
does not have moments.
We note finally that an analogous behavior can be expected for the interfaces defined by
noisy Mullins equation \cite{wv} in $d = 5$ or by general linear Langevin equations described in 
Refs.\cite{luse,krugg} in $d = 3$.

\begin{acknowledgments}

The author acknowledges helpful discussions with P.Krapivsky.

\end{acknowledgments}

\end{document}